\documentclass[aip,aps,preprint,showpacs]{revtex4}
\usepackage{amssymb}

\usepackage{graphicx}

\usepackage[T1]{fontenc}

\usepackage[american]{babel}

\begin{document}

%\title{Nanostructured thermoelectric Ca$_3$Co$_4$O$_9$ films on Al$_2$O$_3$ ceramic templates}
\title{High-throughput synthesis of thermoelectric Ca$_3$Co$_4$O$_9$ films}

\author{D.~Pravarthana}
\affiliation{Laboratoire CRISMAT, CNRS UMR 6508, ENSICAEN, Normandie Universit\'e, 6 Bd Mar\'echal Juin, F-14050 Caen Cedex 4, France.}
\author{O.I.~Lebedev}
\affiliation{Laboratoire CRISMAT, CNRS UMR 6508, ENSICAEN, Normandie Universit\'e, 6 Bd Mar\'echal Juin, F-14050 Caen Cedex 4, France.}
\author{S.~Hebert}
\affiliation{Laboratoire CRISMAT, CNRS UMR 6508, ENSICAEN, Normandie Universit\'e, 6 Bd Mar\'echal Juin, F-14050 Caen Cedex 4, France.}
\author{D.~Chateigner}
\affiliation{Laboratoire CRISMAT, CNRS UMR 6508, ENSICAEN, Normandie Universit\'e, 6 Bd Mar\'echal Juin, F-14050 Caen Cedex 4, France.}
\author{P.A.~Salvador}
\affiliation{Department of Materials Science and Engineering, Carnegie Mellon University, 5000 Forbes Ave., Pittsburgh, Pennsylvania 15213.}
% PAS I added my middle initial and the street address to last two lines.
\author{W.~Prellier}\thanks{wilfrid.prellier@ensicaen.fr} 
\affiliation{Laboratoire CRISMAT, CNRS UMR 6508, ENSICAEN, Normandie Universit\'e, 6 Bd Mar\'echal Juin, F-14050 Caen Cedex 4.}

\date{\today}

\begin{abstract}
\quad Properties of complex oxide thin films can be tuned over a range of values as a function of mismatch, composition, orientation, and structure. Here, we report a strategy for growing structured epitaxial thermoelectric thin films leading to improved Seebeck coefficient. Instead of using single-crystal sapphire substrates to support epitaxial growth, Ca$_3$Co$_4$O$_9$ films are deposited, using the Pulsed Laser Deposition technique, onto Al$_2$O$_3$ polycrystalline substrates textured by Spark Plasma Sintering. The structural quality of the 2000 \AA~ thin film was investigated by Transmission Electron Microscopy, while the crystallographic orientation of the grains and the epitaxial relationships were determined by Electron Back Scatter Diffraction. The use of a polycrystalline ceramic template leads to structured films that are in good local epitaxial registry. The Seebeck coefficient is about 170 $\mu$V/K at 300 K, a typical value of misfit material with low carrier density. This high-throughput process, called combinatorial substrate epitaxy, appears to facilitate the rational tuning of functional oxide films, opening a route to the epitaxial synthesis of high quality complex oxides.

\end{abstract}


\maketitle

\newpage

%\section{Introduction}
%%%%%%%%%%%%%%%% INTRODUCTION %%%%%%%%%%%%%%%%%%%%%%%%
Functional materials have properties that are sensitive to their environment, such as the external pressure, temperature, or magnetic/electric field. Such materials are desirable for a spectrum of applications and thin films are of interest both because the form is technologically useful and they afford the ability to tune further the functional properties. Among the diverse functional properties, thermoelectricity (the property of a material that converts heat into electricity (and vice versa) through Peltier and Seebeck effects) is of interest because it can be used for refrigeration, cooling microelectronic devices, and providing an energy source from waste heat, and the nano-/micro-structural  features that improve properties are controllable using thin film methods.\cite{thermo}

The quality of thermoelectric materials is quantified by the unit less figure of merit ZT, which is calculated via Eq. \ref{ZT}, where T, S, $\rho$, and $\kappa$ are temperature, thermoelectric power (or Seebeck coefficient), electrical resistivity, and thermal conductivity, respectively. Thus, good thermoelectric materials should have a high Seebeck coefficient, a low resistivity, and a low thermal conductivity:

\begin{equation} \label{ZT}
ZT = \frac{S^{2}T}{\rho\kappa}
\end{equation}

For thermoelectric applications, it is consequently required to engineer material exhibiting low electric resistivity as well as thermal conductivity. The electron contribution to the thermal conductivity is proportional to electrical conductivity. Consequently, in most of the materials, a high thermal conductivity is often associated with a high electrical conductivity. In order to enhance thermoelectric figure of merit in such materials, it is necessary to reduce the phonon contribution to the thermal conductivity. A possible way to achieve this, is to increase the phonon scattering at grain boundaries or internal interfaces. For these reasons, several nanostructured thermoelectric materials have been developed, including as nanowires, superlattices, and nanocomposites.\cite{ohta,nano}

Among potential oxide thermoelectric materials, the layered cobalt oxide Ca$_3$Co$_4$O$_9$ is particularly interesting. Ca$_3$Co$_4$O$_9$ can be denoted as $[Ca_2CoO_3]^{RS}[CoO_2]_{1.62}$ to highlight the two layers that constitute the misfit structure and the ratio of the differing $b$-axis parameters.\cite{masset} The two layers are a Ca$_2$CoO$_3$ rock salt-like layer (RS) and a CoO$_2$ cadmium iodide-like layer. These two layers have similar $a$ and $c$ lattice parameters with differing $b$ lattice parameters, where the ratio of the b parameters for the Ca$_2$CoO$_3$ layer to CoO$_2$ layer is 1.62. The Ca$_2$CoO$_3$ layer is a distorted rock salt layer with in-plane Co-O distances of 2.28 \AA and out-of-plane Co-O distances of 1.82 and 1.89 \AA. The CoO$_2$ layer has edge-sharing CoO$_6$ octahedra with Co-O distances of 1.86 and 1.96 \AA.\cite{masset}

Due to the anisotropic nature of Ca$_3$Co$_4$O$_9$, both grain size and grain orientation (i.e., texture) must be controlled to tailor thermoelectric properties.\cite{hiromichi}  There are several ways to process ceramics that afford control over microstructure and texture relevant to given applications.\cite{70 ways} Spark Plasma Sintering (SPS) is an emerging processing technique that can be used to densify fully different kinds of materials. SPS process is a pressure-assisted pulsed current sintering process in which densification is highly promoted at lower temperatures over conventional processes. SPS usually leads to a highly dense ceramic with fine control of grain structures.\cite{SPS} Lamellar ceramics have been processed using SPS and they had much enhanced thermoelectric properties due to texture developed in the SPS process.\cite{noudem,noudem2}

Thin film processing methods also can be used to fabricate dense, textured, layered compounds, and they provide control over other microstructural features, such as grain size. It has already been suggested that improved thermoelectric properties could be achieved in epitaxial Ca$_3$Co$_4$O$_9$ films on single-crystal Al$_2$O$_3$ substrates.\cite{sakai,eng} However, engineering grain size in thin films of complex layered phases is difficult when depositing on macroscopic single crystals.\cite{CCOfilms}  On the other hand, if films are deposited on polycrystalline substrates and local epitaxy is achieved, then the grain size may be controlled in thin films via template effects. Such high quality local epitaxial growth of structurally dissimilar materials has recently been shown for the growth of TiO$_2$ films on perovskites.\cite{burbure,zhang} Deposition of Ca$_3$Co$_4$O$_9$  on polycrystalline Al$_2$O$_3$ substrates has also recently been reported.\cite{Kang} In this article, we explore more closely the local epitaxy of Ca$_3$Co$_4$O$_9$ films (deposited by the Pulsed Laser Deposition (PLD) technique) on polycrystalline Al$_2$O$_3$ substrates, textured using SPS. When the grains of polycrystalline Al$_2$O$_3$, whose size is carefully controlled, act as seeds for localized epitaxial growth, one expects a large range of orientations, and these must be characterized with local structural probes. Here we use electron backscatter diffraction (EBSD) as the local probe, similar to the recent reports.\cite{burbure,zhang} The coalescence of the films is also expected to be different on polycrystalline ceramics than on single crystal substrates, and may depend on growth conditions and substrate parameters. Finally, the deposition of layered films having local epitaxial registry on simple polycrystalline substrates will not only allow the engineering in thermoelectric films, it also opens the path to
 grow complex films with enhanced electronic properties.

%%%%%%%%%%%%%%%% EXPERIMENTAL %%%%%%%%%%%%%%%%%%%%%%
%\section{Experimental Section}
%%%%%%%%%%%%%%%% INTRODUC
Undoped commercially-available $\alpha$-Al$_2$O$_3$ (ALO) powders were sintered using Spark Plasma Sintering (SPS), as describe elsewhere.\cite{pravat} Briefly, an appropriate amount of powder was pressed under 100 MPa at 1700$^{\circ}$C.(see note below) The crystalline phase was confirmed to be corundum using conventional x-ray diffraction. The sintered samples were polished successively up to a 3-4 nm roughness for EBSD characterization and subsequent film growth. They were: mechanically polished down to 10 $\mu m$ roughness, then, polished with diamond paste of 3 $\mu m$ and  1 $\mu m$, respectively, for about 2 minutes, resulting in a mirror-like surface. Finally, the polished surface was etched in 5 \% HF:HNO$_3$ solution to remove surface contaminants and release strain due to polishing.\cite{polishing} 

The deposition of the Ca$_3$Co$_4$O$_9$ (CCO) films was performed at 650 $^{\circ}$C under 0.2 mbar of oxygen pressure using the pulsed laser deposition technique ($\lambda = 248 nm$). A laser energy of 2 J/cm$^2$ and repetition rate of 3 Hz were typically used. The films with thickness ranging from 500 to 5000 \AA were grown. The composition of the films was checked by Energy Dispersive Analysis and found to be close to the target within the experimental error. Structural and microstructural characterizations of the ceramics and the films were carried out using Electron BackScatter Diffraction (EBSD) and Transmission Electron Microscopy (TEM). For EBSD analysis, the samples were mounted with a 70$^{\circ}$-tilt angle from horizontal in a scanning electron microscope (SEM) operated at 20 kV. Note that the thickness of the films (>5000 \AA) is well above the EBSD probing depth (20 nm for an electron acceleration voltage of 20 kV)\cite{resolution} to make sure that the EBSD analysis comes only from the film. The acquired Kikuchi patterns were indexed automatically by the EDAX Orientation Imaging Microscopy (OIM$^{TM}$) software (v.6) after two iterations removing points less than 2 $\mu m$ and a grain tolerance less than 3$^{\circ}$. The Electron Back Scatter Diffraction (EBSD) experiments were performed on both the surface of the Ca$_3$Co$_4$O$_9$ films, as well as the polycrystalline Al$_2$O$_3$ substrate to understand the nature of the surrounding grains, its influence on the orientation of the Ca$_3$Co$_4$O$_9$ film and their epitaxial relationships. 

TEM investigations were carried out using a FEI Tecnai G2 30 UT microscope operated at 300 kV (point resolution 1.7 \AA). Cross-section samples were cut parallel to the Ca$_3$Co$_4$O$_9$/Al$_2$O$_3$ interface plane, mechanically polished to a thickness of about 15 $\mu$m, followed by Ar+ ion beam milling under grazing incidence with respect to the surface. A soft regime of ion milling was used to prevent any possible artifacts arising from TEM specimen preparation. Image simulations were made with CrystalKit and MacTempas software. The structure and  epitaxial relationship of the film on the polycrystalline substrates was investigated in detail. Finally, the Seebeck coefficient ($S$) and the resistivity ($\rho$) of the films were measured as a function of temperature ($T$) in Physical Properties Measurements System (PPMS, Quantum Design), using a standard four probe ($\rho$) and steady state techniques ($S$).

%\section{Results and Discussion}
%%%%%%%%%%%%%%%% INTRODUC
%%%%%%%%%%%%%%%% RESULTS %%%%%%%%%%%%%%%%%%%%%

%%%%%%%%%%%%%%%% FIGURE 1 %%%%%%%%%%%%%%%%%%%%%

The grain orientation map of the Al$_2$O$_3$ ceramic substrate is shown in Figure 1a as an inverse pole figure, along with the appropriate color-coded stereographic triangle for a triganal system. For Al$_2$O$_3$ substrate, the patterns were indexed best in the trigonal system using a=4.785 \AA\ and c=12.991 \AA\ (ICSD file No.169720). In this plot, grains having low confidence index (below 0.1) have not been considered. The average image quality was IQ=1224, which confirms the high crystalline nature of grains and the near surface region. The grain size (mean area equivalent diameter) was about 50 $\mu$m, but exhibited a clear bimodal grain size distribution, with an isometric grains on the order of 100 $\mu$m and more isometric grains on the order of 10 $\mu$m. The misorientation distribution values across the grain boundaries ranged from 15 to 120$^{\circ}$, and the relative fraction of boundaries in different ranges of misorientation angle are given in Figure 1c. The majority of the grain boundaries are high angle boundaries, within 30$^{\circ}$ of 60$^{\circ}$. That the ceramic exhibits large misorientation angles across grain boundaries and anisometric (elongated) grain growth is probably related to the trigonal symmetry of  Al$_2$O$_3$. More details of the EBSD analysis of such SPS fabricated Al$_2$O$_3$ ceramics can be found elsewhere.\cite{pravat} The crystalline quality of the Al$_2$O$_3$ grains and grain boundaries, which serve as templates for the oriented growth of the Ca$_3$Co$_4$O$_9$ films, were investigated by transmission electron microscopy (TEM). A typical high-resolution TEM image of a grain boundary is shown in Figure 1d. The two grain are highly crystalline and the boundary is sharp and well-defined.

%\textbf{You should comment on what is known about 60 $^{\circ}$ twin boundaries in Al2o3- do these explain the cluster around 60? Are the fraction length fractions or number fractions? You should reference the symmetry driven micro structural features, this must be well known.}
%\textbf{PAS Mar 06, I modified the text in the figure order I thought made sense. Also, did you just ignore points or is the image cleaned up (partitioned and cleaned)}

%%%%%%%%%%%%%%%% FIGURE 2 %%%%%%%%%%%%%%%%%%%%%

After PLD growth of the Ca$_3$Co$_4$O$_9$ film, similar analyses were carried out on the surface of the film. EBSD patterns from the grains marked as G1, G2, and G3 in Figure 1a are shown in Figure 2, both from the substrate before deposition (Figure 2a,c,e) and the film after deposition (Figure 2b,d,f). Clear contrast in the EBSD patterns are observed for all cases, though the patterns are slightly more diffuse for the film (as reported previously),\cite{burbure,zhang} arising from internal strains in the film introduced by the epitaxial growth on the underlying grain.\cite{strain} It should be noted that several film patterns were registered for each grain, and the patterns were similar as long as they were on the same underlying substrate grain. Using the automated procedure, the orientation of grains 1, 2, and 3 of Al$_2$O$_3$ can be assigned from the detected zone axis shown in the figure as [$01\bar{1}\bar{1}$], [$\bar{2}021$], and [$\bar{1}\bar{1}20$] respectively. The three-fold symmetry axis confirms the trigonal crystal symmetry.\cite{resolution} For Ca$_3$Co$_4$O$_9$ film, the indexing is more complicated, since the Ca$_3$Co$_4$O$_9$ film is a misfit-layered oxide and consists in two monoclinic subsystems with identical $a$, $c$, and $\beta$ parameters, but different $b$ parameters.\cite{masset} Thus, it is expected, that the backscatter diffraction pattern of the Ca$_3$Co$_4$O$_9$ would be a superposition of those of two subsystems and difficult to be interpreted.\cite{tani} For this reason, we adopted simple monoclinic lattice parameters (a=4.834 \AA, b=4.558 \AA and c=10.844 \AA and $\alpha, \beta, \gamma = 90, 98.141, 90^{\circ}$, respectively) for the analysis. Note that the $b_2$ parameter for the rock salt-type sub-system was used as $b$ axis length, as suggested previously.\cite{tani} The Kikuchi images can be indexed for the film, and confirmed that the Ca$_3$Co$_4$O$_9$ film is of high structural quality. The series of Ca$_3$Co$_4$O$_9$ grains, G1, G2 and G3 are thus indexed as [$407$], [$42\,15$], and [$11\,\bar{3}4$], respectively.\cite{babakishi} (The orientation relative to the crystal, obtained from Kikuchi patterns, can also be represented with three Euler angles and these are given for each image in the caption of Figure 2.) 

We quantitatively investigated the crystallographic orientation of the films relative to the ceramic substrate by comparing the location of the ${001}$ pole for both film and ceramic. These pole figures are given in Fig. 3a for the substrate grains and 3b for the film grains. Values registered from 56 randomly selected grains are shown, and one observes that the poles are well-distributed in angular space for both materials. When on compares the relative location of the ${001}$ pole for a given film grain on a specific substrate grain, we determine an average of 5$^\circ$ between the c-axis of Al$_2$O$_3$ and the c-axis of Ca$_3$Co$_4$O$_9$. This overlap between the pole points confirms that the [0001] and [001] directions of Al$_2$O$_3$ and Ca$_3$Co$_4$O$_9$ films, respectively are essentially parallel. The rotation of 5$^\circ$ between the film and the substrate results from a combination of experimental uncertainty in the orientation measurements and misalignment of the sample during the two measurements, as observed previously.\cite{zhang} Essentially, this argues that local texture develops in our samples not as a preference of the c-axis to orient itself normal to the growth direction, as observed previously,\cite{eng} but owing to a preference of the epitaxial arrangement of the two crystal lattices upon one another during growth- regardless of the absolute orientation of the grain. This is similar to what has been observed for TiO$_2$ growth on perovskite BiFeO$_3$, another case of non-isostructural epitaxial growth on high-miller index plane polycrystalline surfaces.\cite{burbure,zhang} The current case of Ca$_3$Co$_4$O$_9$ growth is of special interest because the crystal structure is much more complex than anatase and rutile TiO$_2$, yet the local epitaxial growth is maintained on high-index surfaces.

Figure 4a shows a cross-sectional low-magnification TEM image of the interface between a 500 nm thick Ca$_3$Co$_4$O$_9$ film and the Al$_2$O$_3$ substrate. A grain boundary exists in the center portion of the film (marked with a vertical arrow), and the two grains are marked G1 and G2 (these are not the same as the G1 and G2 grains in Fig. 1). The surface of the Ca$_3$Co$_4$O$_9$ film is not flat, and this is particularly evident in Fig. 4a, where a ''wave-like'' surface is observed for the film. This waviness correlates with the different tilting of the c-axis with respect to the substrate surface on the two grains in Al$_2$O$_3$, as well as with the surface roughness of the near the grain boundary. In other words, the roughness comes both from differences in the local orientation of the film and the underlying roughness of the polished polycrystal. Interestingly, each film grain is growing on each substrate grain, suggesting a local epitaxy grain over grain. The corresponding SAED pattern (covering the entire region of G1 and G2) is shown in an inset, and correlates to the superposition of the electron diffraction pattern of the substrate (white rectangle) and of the film containing two rotated c-axis grains corresponding to the tilted $c$-oriented grains observed in Fig. 4a. The presence of an internal Al$_2$O$_3$ reference in ED pattern allows one to determine the $c$-axis lattice parameter of Ca$_3$Co$_4$O$_9$ material with high accuracy; the determined value of $c$=10.78 \AA\ is in reasonable agreement with the 10.83 \AA\ value reported in single crystals and thin films.\cite{masset,CCOfilms,BiCCO,CoO2}

%\textbf{PAS Mar 06 I am not sure about the SAED... are there 4 grains- two substrate and two film, or just three : one substrate and two films. If it is the first, great, and clarify. If it is the second, then why are there two film grains... it needs to be explained.}

%From the Electron Diffraction (ED) pattern (Fig.3a) some epitaxial relationships can be determine as following: (011)$_{ALO}$//(110)$_{CCO}$; [5-44]$_{ALO}$//[001]$_{CCO}$.

An HRTEM image of the Ca$_3$Co$_4$O$_9$ film is shown in Fig. 4b, recorded along [110] zone axis of Al$_2$O$_3$. Bright contrast bands are clearly visible along $c$-axis direction, which arise from a small tilting out of the exact [110] zone axis. The image area also contains CoO$_2$ stacking faults, which are typical for Ca$_3$Co$_4$O$_9$ film as previously reported.\cite{CoO2} The image simulation of Ca$_3$Co$_4$O$_9$ is given as an inset to Fig.3c, and is based on the monoclinic Ca$_3$Co$_4$O$_9$ structure (SG C2/m with a=4.8376 \AA, b=4.5565 \AA, c=10.833 \AA, and $\beta=98.06^{\circ}$). The image simulation shows good agreement with the experimental image.\cite{masset} The epitaxial growth of Ca$_3$Co$_4$O$_9$ on Al$_2$O$_3$ substrate is clear in the cross-sectional HRTEM image given in Fig. 4d for the film/ceramic interface, taken along the [011] Al$_2$O$_3$ zone. The Fourier transform (FT) pattern given in the lower inset in Fig. 4d exhibits spots from both [011] Al$_2$O$_3$ and spots from the grain of Ca$_3$Co$_4$O$_9$ film. The FT pattern given in the upper inset of Fig. 4d is from only the Ca$_3$Co$_4$O$_9$ grain, and is indexed as a [201]$_{CCO}$ oriented grain. Therefore, some epitaxial relationships can be expressed for this grain: [011]$_{ALO}$//[201]$_{CCO}$; [100]$_{ALO}$//[$11\bar{2}$]$_{CCO}$; [$1\bar{2}2$]$_{ALO}$//[$\bar{1}12$]$_{CCO}$. It is important to notice that no secondary phase or amorphous layer were observed at the interface. This is very different from the previous TEM analysis of Ca$_3$Co$_4$O$_9$ thin films grown on single crystal sapphire (001)-oriented, glass, or (100)-silicon, where the presence of both secondary phases and low-crystallinity structures at the interface were reported.\cite{tem1,tem2,tem3} In our present study, the absence of an amorphous layer close to the interface clearly confirmed the good texture and the epitaxial nature of the Ca$_3$Co$_4$O$_9$ thin film with respect to each grain, and will be discussed hereafter with respect to the thermoelectric properties. It should be clear that the Ca$_3$Co$_4$O$_9$ epitaxial film growth was achieved without the \textit{classical} substrate-induced epitaxy arising from interactions with low-index surfaces of single-crystal substrates, which is very different that previous reports.\cite{Kang,CoO2}

The temperature-dependence of the Seebeck coefficient and the resistivity are shown on Figure 5. Contrary to the single crystals or polycrystals of Ca$_3$Co$_4$O$_9$,\cite{masset,vac} the resistivity presents here a localized behaviour, with values of resistivity close to 100 $\mu\Omega$.cm at 300 K, and a local activation energy d(ln$\rho$/d(1/T) increasing from 50 meV at low temperature to 300 meV at 300 K. The film resistivity increases dramatically below 150 K (see insert of figure 5), and the activation energy is calculated to be 25.2 meV. Note that the resitivity becomes too high at low temperature for an accurate measurement of the thermoelectric power. From 150 K to 300 K, the film presents an almost constant value of Seebeck close to 170 $\mu$V/K. As seen in the bulk, the positive value indicates hole-like carriers, and the shape of $S(T)$ is typical of Ca$_3$Co$_4$O$_9$, with a plateau observed for T > 100-200K, which follows a steep increase of $S$ at low temperature (which can not be observed here due to the large value of resistivity). However, the value of S (close to 170$\mu$V/K at 300 K) is slightly larger than for the bulk crystals (120 $\mu$V/K),\cite{masset} leading to a power factor, S$^2/\rho$, calculated to be 0.025 mWm$^{-1}$K$^{-2}$, a value similar to previous thin films also made by laser ablation but deposited on single crystals: 0.05 mWm$^{-1}$K$^{-2}$ \cite{eng} and 0.03 mWm$^{-1}$K$^{-2}$.\cite{Kang} These power factors are smaller that the highest power factor reported for bulk samples, which are ~0.5 mWm$^{-1}$K$^{-2}$.\cite{dalton} In bulk misfits, an increase of Seebeck coefficient, associated to a more localized behaviour has also been observed. It can be due to a reduction of Co$^{4+}$ content, which can be in fact induced by oxygen non-stoichiometry and/or a modification of the misfit ratio $b1/b2$. A relatively large enhancement of the Seebeck coefficient reaching 178 $\mu$V/K at 300 K has also been observed in bulk Ca$_3$Co$_4$O$_9$ samples when prepared using high magnetic field sintering compared with conventional sintering and SPS techniques, and attribute to a better control of the oxygen non-stoichiometry.\cite{huang} In the case of thin films, while oxygen content is definitively crucial,\cite{klie} substrate-induced strain can also be an important issue for the thermopower datas.\cite{CoO2} A detailed investigation of the interplay between the substrate and the film which shows values larger than 125 $\mu$V/K, has also been reported using different kinds of substrates, but most of the films present not only defects at interface but also CoO$_2$ stacking faults.\cite{CoO2} Here, the structured films do not present amorphous layer at the interface, and the number of stacking faults is small, which confirms their structural quality, and the local epitaxy. Hall effect measurements performed at 300 K on films prepared under similar conditions display a carrier density of 10$^{18}$ cm$^{-3}$ suggesting that the enhancement of Seebeck coefficient is mostly due to this small carrier concentration, in agreement with our resistivity values.

%\section{Conclusion}
%%%%%%%%%%%%%%%% INTRODUC
%%%%%%%%%%%%%%%% CONCLUSION %%%%%%%%%%%%%%%%%%%%%%

%Films were grown on large grain size alumina substrate to minimize the strain-induced by grain boundaries.

In conclusion, high-throughput synthesis of Ca$_3$Co$_4$O$_9$ thin films was realized on textured Al$_2$O$_3$ ceramic templates prepared by spark plasma sintering. Microstructure analysis reveals a good local epitaxy of the film over the grain boundaries. Electrical measurements show a maximum Seebeck coefficient of 170 $\mu$V/K at 300 K, slightly higher than the bulk value, but typical of misfit materials. The utilization of textured ceramic substrates appears promising for meticulous control of the structure and orientation of the film growth through the oriented-grain strain engineering. In a broad perspective, this high-throughput synthesis called combinatorial substrate epitaxy, can be expanded to nanostructure functionalized oxide films.

We thank L. Gouleuf and J. Lecourt for technical support. D.P. is support by a PhD fellowship included in the Erasms Mundus Project IDS-FunMat. Partial support of the french Agence Nationale de la Recherche (ANR), through the program Investissements d'Avenir (ANR-10-LABX-09-01), LabEx EMC3 is also acknowledged.

\newpage
Figures Captions

Figure 1: (a) : 300x300 $\mu m^2$ Inverse Pole Figure (IPF) of Al$_2$O$_3$ substrate which gives the crystallographic orientations orthogonal to sample surface. Regions with low confidence in the indexing (<0.1) have been removed. The orientation map included some boundaries that have been highlighted. White lines: subgrain boundaries with the very small misorientation angles (15-30$^\circ$); black lines: subgrain boundaries with the small misorientation angles (30-60$^\circ$); red lines: grain boundaries with the large misorientation angles (60-90$^{\circ}$); green lines: grain boundaries with the large misorientation angles (90-120$^\circ$). The representation of the colour code (c) used to identify the crystallographic orientations on standard stereographic projection is also indicated (red: [$0001$]; blue: [$1\bar{1}00$]; green: [$0\bar{1}10$]). Note that each colour represents one single grain. The number of fractions of boundaries with different misorientation angles is reported in the table (c). (d): HRTEM bright field image of Al$_2$O$_3$ grain boundary (GB) as marked by arrow. 

Figure 2: EBSD patterns for the Al$_2$O$_3$ substrate and Ca$_3$Co$_4$O$_9$ films for several grains (G1, G2 and G3, see Fig.1). (a), (c) and (d) refer to the Al$_2$O$_3$ substrate. (b),(d) and (f) refer to Ca$_3$Co$_4$O$_9$ films. Substrate (film) Euler angles are: G1 - 235.6 (296.1), 105.9 (19.8), 9.9 (77.3); G2 - 101.3 (99.9), 70.2 (170.8), 283.2 (296.9); G3 - 86 (178.1), 72.7 (104.8), 285.9 (286.9).

Figure 3: Grain texture of film and ceramic represents in <001> pole figure. (a) Al$_2$O$_3$ ceramic substrate before deposition, (b) Ca$_3$Co$_4$O$_9$ film. The angle of grains labelled A, B and C with respect to the center of the pole is 79.21, 29.82 and 9.2$^\circ$, for Al$_2$O$_3$ and 73.12, 77.57 and 22.94$^\circ$ for Ca$_3$Co$_4$O$_9$. The center of the figure is the normal direction. Concentric circles are scaled every 15$^{\circ}$.

Figure 4: (a) Low magnification cross-section bright-field TEM image of Ca$_3$Co$_4$O$_9$ film grown on Al$_2$O$_3$ polycrystalline substrate showing the interface region between the first and second c-axis orientation of Ca$_3$Co$_4$O$_9$ film. The corresponding SAED pattern is given as insert, and is a superimposition of film and substrate (few grains). The GB between two grains of Al$_2$O$_3$ is marked with white arrow. (b) HRTEM image of c-axis oriented Ca$_3$Co$_4$O$_9$ film along [110] zone axis and (c) enlargement of selected in (b) with white box area together with simulated image ($\Delta f=-70 nm, t= 6 nm$) given as insert. The CoO$_2$ stacking faults are indicated by white arrows. (d) HRTEM image of Al$_2$O$_3$/Ca$_3$Co$_4$O$_9$ interface and corresponding FT patterns taken from interface (bottom) and Ca$_3$Co$_4$O$_9$ grain only (top one). Note the absence of amorphous layer at the interface or the presence of secondary phase.

Figure 5: Temperature-dependence of the Seebeck coefficient and resistivity (insert).

\end{document}